\documentstyle[aps,preprint]{revtex}

\newcommand{\rr}{\chi}

\begin{document}
\draft
\tightenlines
\preprint{SNUTP-98-012}

\title{Quantum Inflaton Dynamics}

\author{Dongsu Bak, $^a$\footnote{Electronic address: dsbak@mach.scu.ac.kr}
Sang Pyo Kim$^b$\footnote{Electronic address: sangkim@knusun1.kunsan.ac.kr},
Sung Ku Kim, $^c$\footnote{Electronic address: skkim@theory.ewha.ac.kr},
Kwang-Sup Soh, $^d$\footnote{Electronic address: kssoh@phya.snu.ac.kr},
and Jae Hyung Yee $^e$\footnote{Electronic address: jhyee@phya.yonsei.ac.kr}}

\address{$^a$ Department of Physics,
University of Seoul,
Seoul 130-743, Korea \\
$^b$ Department of Physics,
Kunsan National University,
Kunsan 573-701, Korea \\
$^c$ Department of Physics,
Ewha Womans University,
Seoul 120-750, Korea\\
$^d$ Department of Physics Education,
Seoul National University,
Seoul 151-742, Korea\\
$^e$ Department of Physics,
Yonsei University,
Seoul 120-749, Korea}
\date{\today}

\maketitle
\begin{abstract}
We show that the quantum dynamics of a real scalar field for a
large class of potentials in the symmetric Gaussian state,
where the nonperturbative quantum contributions are taken into
account, can be described equivalently by a two-dimensional nonlinear dynamical
system with a definite angular momentum (U(1) charge of a complex theory).
It is found that the Gaussian state
with a nearly minimal uncertainty and a large quantum fluctuation, as an
initial condition, naturally explains the most of the essential features
of the early stage of the inflationary Universe.
\end{abstract}
\pacs{PACS number(s): 98.80.Cq; 04.62.+v}

The standard cosmological model combined with  particle physics at high
energy (temperature) is plagued with the monopole,  the horizon,
the flatness problems, and so on. A major breakthrough to solve most of these
problems was found in the inflation paradigm invented by Guth
\cite{guth}. (For review and references, see
\cite{linde}.) His simple but attractive idea, the so-called old
inflation scenario, is that an inflaton (homogeneous
scalar field) undergoes a first order phase transition from a
symmetric vacuum as the Universe expands and thereby the
temperature drops, and the energy density of inflaton captured in the
false vacuum drives an exponential expansion of the de Sitter phase.
However, this scenario has a major defect, the graceful exit
problem \cite{hawking}. The new inflation scenario
\cite{linde2} based on a second
order phase transition overcomes the graceful exit problem, but
raises another problem involving the  fine tuning of coupling constants.
It is also possible to solve the graceful exit problem in the
context of Jordan-Brans-Dicke theory \cite{la}.
The chaotic inflation scenario introduced by Linde
\cite{linde3} does not make use of phase transitions but rather
investigate various initial distributions of the inflaton that
lead to inflation. Furthermore, this scenario is model-independent
in the sense that one can obtain a significant inflation for a
large class of potentials.

All these scenarios are based on the classical gravity of the
Friedmann equation and the scalar field equation on the
Friedmann-Robertson-Walker (FRW) universe, assuming its validity
even at the very early Universe. However, quantum effects of
matter fields such as quantum fluctuations are expected to play a
significant role in this regime, though quantum gravity effects
are still negligible. The proper description of cosmological model
should be in terms of the semiclassical gravity of the
semiclassical Friedmann equation with quantized matter fields as
far as inflation is concerned.

In this Letter, we pursue the  quantum dynamics of a
real-valued inflaton (homogeneous real
scalar field) for a  large class  of potentials in the FRW
Universe. In the  Schr\"{o}dinger picture, the scalar inflaton
theory is approximately but quite accurately described by a
symmetric Gaussian state for an extremal configuration
of a bosonic condensate of  scalar particles.
The  symmetric Gaussian state can be explicitly found
by using a nonperturbative method of Ref. \cite{kim} or
a direct Gaussian  wave function as of Refs. \cite{pi,yee}.
The resulting equations are two-dimensional coupled nonlinear equations
where the dynamical variables are
the dispersion of inflaton and the scale factor
of the FRW Universe.

We shall begin with a spatially flat FRW
metric\footnote{We shall use units system  $c = \hbar = 1$ and $m_P^2 = {1 \over G}$
and follow the notations in Ref. \cite{linde3}, in which the Planck length
and time are $l_P = t_P = {1 \over m_P}$, $\phi$ and $a$ have
the dimension $m_P$ and ${1 \over m_P}$, respectively, and $\lambda_{2n}$
is dimensionless.}
\begin{equation}
ds^2= -dt^2+a^2(t) d{\bf x}^2,
\end{equation}
and consider a real homogeneous and isotropic inflaton described by the Hamiltonian
\begin{equation}
\label{e02}
H (t) = {{p^2} \over {2 a^3 (t)}} + a^3 (t) \Biggl({{m^2} \over {2}}\phi^2
+ {{m_P^4 \lambda_{2n}} \over {(2n) }} \Bigl({\phi \over m_P}\Bigr)^{2n} \Biggr) ,
\end{equation}
where $p=a^3 \dot{\phi}$ is the momentum conjugate to the real inflaton field.
The classical equation of motion for the inflaton field reads
\begin{equation}
\ddot{\phi} + 3 \Bigl({{\dot{a}} \over {a}} \Bigr) \dot{\phi} + m^2 \phi +
m_P^2 \lambda_{2n} \Bigl({\phi \over m_P} \Bigr)^{2n-2} \phi = 0 ,
\label{class eq}
\end{equation}
while the the scale factor $a(t)$ is governed by  the Friedmann
equation
\begin{equation}
\Bigl({{\dot{a}(t)} \over {a (t)}} \Bigr)^2 = {{8 \pi} \over {3 a^3 m_P^2}}
H(t) .
\label{class fr eq}
\end{equation}
In this classical context, it is known that the inflation may indeed occur
but the unnatural fine tuning of the initial data
is, more or less, needed as will be seen. But before dealing with the classical
gravity, we shall first analyze the quantum theory and obtain the classical theory
as its limit.

Upon quantization of the real inflaton, the Schr\"odinger equation
for quantum field
\begin{equation}
i {{\partial} \over {\partial t}} \Psi (\phi, t)
= \hat{ H} (t) \Psi (\phi, t) ,
\label{sch eq}
\end{equation}
determines the time evolution of the inflaton, whereas
the Friedmann equation at its semiclassical level
\begin{equation}
\Bigl({{\dot{a}(t)} \over {a (t)}} \Bigr)^2 = {{8 \pi} \over {3 a^3 m_P^2}}
\langle \hat{H}\rangle ,
\label{sem fr eq}
\end{equation}
describes the evolution of the Universe.

There is a rather  standard method  to find an approximate  Gaussian
state that extremizes the energy (see e.g. Refs. \cite{kim,pi,yee}).
Although one may use a more generic Gaussian wave function, we
shall use a symmetric Gaussian state \cite{pi}
\begin{equation}
\Psi(\phi, t) = {1 \over (2 \pi \rr^2 (t))^{1/4}}
\exp\Bigl[- \Bigl({1 \over 4 \rr^2 (t)} - i {\pi_{\rr} (t)
\over 2 \rr(t)} \Bigr)\phi^2 \Bigr],
\label{gaussian form}
\end{equation}
as our trial state. The real functions $\chi$ and $\pi_{\chi}$ are
time-dependent parameters, whose time-dependence will be
determined below by the Dirac action principle.
We are going to extremize the effective action
\begin{equation}
I_{\rm eff}=\int dt \langle \Psi|[i\partial_t -\hat{\bf H}(t)]|\Psi \rangle ,
\label{action}
\end{equation}
which is the Dirac action except for its wave function is now
limited by the Gaussian form in Eq. (\ref{gaussian form}).
By a straightforward computation, one finds that the effective
action is given by
\begin{equation}
I_{\rm eff}=\int dt [\pi_\rr \dot\rr -H_{\rm eff}(\pi_\rr,\rr)] ,
\label{action1}
\end{equation}
where
\begin{eqnarray}
\label{e16}
H_{\rm eff} (\pi_\rr,  \rr) =
{\pi^2_{\rr}  \over 2 a^3} + a^3 V_{\rm eff} (\rr) ,
\label{eff ham}
\end{eqnarray}
with the effective potential,
\begin{eqnarray}
V_{\rm eff} (\rr) =  {1 \over 8 a^6\rr^2} +
{{m^2} \over {2}} \rr^2 +
{m_P^4 \lambda^Q_{2n} \over (2n)}
\Bigl({\rr \over m_P} \Bigr)^{2n},
\label{e17}
\end{eqnarray}
where $\lambda_{2n}^Q = {(2n)!\over 2^n n!} \lambda_{2n}$ is an
effective coupling constant.
Upon extremization of the effective action (\ref{action1}),
one finds that our system is described by the Hamilton's equations
\begin{eqnarray}
a^3 \dot{\rr} &=&\  \pi_\rr ,
\nonumber\\
\dot{\pi_\rr}\  &=& - a^3 {{\partial} \over {\partial \rr}}
V_{\rm eff} (\rr)
\nonumber\\
&=& {{1} \over {4 a^3 \rr^3}}
-  a^3 \Biggl( {m^2} \rr +
m_P^3 \lambda^Q_{2n}
\Bigl({\rr \over m_P} \Bigr)^{2n-1} \Biggr) ,
\label{ham eq}
\end{eqnarray}
The semiclassical Friedmann equation becomes
\begin{equation}
\Bigl({{\dot{a}(t)} \over {a (t)}} \Bigr)^2 = {{8 \pi} \over {3 a^3 m_P^2}}
H_{\rm eff} .
\label{sem fr eq2}
\end{equation}

One can also understand the physics of the above system
from a different point of view using the method of Ref. \cite{kim},
which is based on the technique of
solving the Schr\"{o}dinger equation for the time-dependent
Hamiltonian system \cite{lewis}. One introduces the
annihilation operator of a Fock space \cite{kim2} redefined
as dimensionless quantity,
\begin{equation}
\hat{A} =  {{\varphi}^* (t) \over m_P}  \hat{p} - m_P a^3 (t) \dot{\varphi}^* (t)
 \hat{\phi}, ~ \hat{A}^{\dagger} = {\rm h. c},
\label{op}
\end{equation}
such that
\begin{equation}
\Bigl[\hat{A}, \hat{A}^{\dagger} \Bigr] = i a^3 (\varphi^*
\dot{\varphi} - \dot{\varphi}^* \varphi) = 1,
\label{commutation}
\end{equation}
and $\varphi (t)$ in Eq. (\ref{op}) is a complex scalar variable
with the same dimension as the classical field $\phi$.
One then expands the Hamiltonian in terms
of $\hat{A}$ and $\hat{A}^{\dagger}$, and truncates it up to the
quadratic terms $\hat{H}_{(2)}$ for an approximation. The requirement that
$\hat{A}$ and $\hat{A}^{\dagger}$ should be the solutions of the
Liouville-Neumann equation,
\begin{equation}
i {{\partial \hat{A}} \over {\partial t}} + \left[ \hat{A},
\hat{H}_{(2)} \right] = 0 ,
\end{equation}
leads to the equation of motion for the $\varphi$-field
\begin{equation}
\ddot{\varphi}  + 3
\Bigl({{\dot{a}} \over {a}} \Bigr) \dot{\varphi} + m^2  \varphi +
m_P^2 \lambda^Q_{2n} \Bigl({{\varphi^*  \varphi}
\over {m_P^2}} \Bigr)^{n-1} \varphi = 0 .
\label{mean}
\end{equation}
If one sets $\varphi = \rr e^{-i \theta}$,  Eq. (\ref{mean})
reduces to Eq. (\ref{ham eq}) with the constraint
(\ref{commutation}) expressed as
\begin{equation}
Q \equiv 2 a^3 \rr^2 \dot{\theta} = 1 .
\label{charge}
\end{equation}
We have thus found that the quantum dynamics of the real inflaton
theory is equivalently described by the two-dimensional nonlinear
dynamical system.
The expectation value of the Hamiltonian gives rise to the
effective Hamiltonian
\begin{equation}
H_{\rm eff} (t) = \langle \hat{H} (t) \rangle = a^3 \Biggl\{
{1 \over 2}\dot{\varphi}^* \dot{\varphi} + {{m^2} \over
{2}}\varphi^* \varphi
+ {{m_P^4 \lambda_{2n}} \over {(2n) }}
\Bigl({\varphi^* \varphi \over m_P^2}\Bigr)^{n} \Biggr\} .
\end{equation}
It should be noted that this complex system has a U(1)
symmetry under the transformation $\varphi \rightarrow e^{i \alpha}
\varphi$, and that the commutation relation $\Bigl[\hat{A} ,
\hat{A}^{\dagger} \Bigr] = 1$ indeed determines the charge
to be unity. In this framework one can easily obtain the higher
order corrections to the Gaussian state (\ref{gaussian form})
\footnote{For the
systematic improvement of the approximation, see
Ref. \cite{kim} and the detailed deviations from the exact results  in this
Gaussian approximation, is dealt with in Ref. \cite{yee}.}.

A few comments are in order. First, we note that the effective
potential (\ref{e17}) comes from the nonperturbative quantum  contributions
and the factor ${{(2n)!} /( {2^n n!}})$ of $\lambda_{2n}^Q$ in
Eqs. (\ref{ham eq}),(\ref{mean}) and (\ref{sem fr eq2}),
actually accounts for the number of symmetric loop diagrams from the higher order
self-interactions and, hence, enhances the effective coupling constants
in the very early Universe when the quantum effects of matter fields
are expected to be important.
For the massive real inflaton without higher order interaction
(i.e. $\lambda_{2n} = 0 $), the wave function (\ref{gaussian form}) determined
in this way  is indeed the exact state of
Eq. (\ref{sch eq}), and the other excited quantum states can be
constructed by acting the creation operators (\ref{op}) \cite{kim2}.
Once the higher order coupling is turned on,
the Gaussian state is no longer exact, but it is also known that it
describes quite accurately the system  even for a strong coupling constant.

We are now able  to find the two asymptotic solutions to Eq. (\ref{ham eq}).
First, we consider the quantum dynamics at the very early Universe.
Eq. (\ref{ham eq}) can be written as a second order equation
\begin{equation}
\ddot{\rr} + 3 \Bigl({{\dot{a}} \over {a}} \Bigr) \dot{\rr}
- {1 \over 4 a^6 \rr^3} +  {m^2} \rr +
m_P^3 \lambda^Q_{2n} \Bigl( {\rr \over m_P} \Bigr)^{2n-1} = 0.
\label{radial eq}
\end{equation}
The symmetric Gaussian state (\ref{gaussian form})
for the inflaton has the dispersions
\begin{eqnarray}
\Delta \phi &=& \sqrt{ \langle \hat{\phi}^2 \rangle
- \langle \hat{\phi} \rangle^2} = \rr,
\nonumber\\
\Delta \pi &=& \sqrt{ \langle \hat{\pi}^2 \rangle
- \langle \hat{\pi} \rangle^2} = a^3 \Bigl(\dot{\rr}^2 + \rr^2
\dot{\rr}^2 \Bigr)^{1/2},
\label{dispersion}
\end{eqnarray}
where $\langle \hat{\pi} \rangle = \langle \hat{\phi} \rangle =
0$.
The uncertainty relation becomes
\begin{equation}
\Delta \phi \Delta \pi = {1 \over 2} \Bigl(1 + 4 \pi_{\rr}^2 \rr^2
\Bigr)^{1/2},
\label{uncertainty}
\end{equation}
where we used Eq. (\ref{charge}).
A nearly minimal uncertainty can be achieved
when $\pi_{\rr} \rr \sim 0$, that is,
either $\pi_{\rr} \sim 0$ or $\rr \sim 0$.
We propose an inflaton's initial quantum
state with $\pi_{\rr} \sim 0$ and a large quantum fluctuation $\rr >
\rr_c$, where
\begin{equation}
{\rr_c  \over m_P} = {\min.} \Biggl\{\Bigl[{1 \over \lambda_{2n}^Q}
\Bigl({m \over m_P} \Bigr)^2 \Bigr]^{1/(2n -2)} ,
 \Bigl[{1 \over 4 \lambda_{2n}^Q}
\Bigl({m \over a_0} \Bigr)^6 \Bigr]^{1/(2n +2)} \Biggr\}.
\end{equation}
Under this initial condition the first term and
the centrifugal potential term in Eq.
(\ref{radial eq}) are small compared with the other two terms,
and Eq. (\ref{radial eq}) reduces to the equation for
classical inflaton when the dispersion $\rr$ is interpreted as $\phi$.
For this large quantum fluctuation,  Eq. (\ref{radial eq})
may be approximated as
\begin{equation}
3 \Bigl({{\dot{a}} \over {a}} \Bigr) \dot{\rr}  +
m_P^3 \lambda^Q_{2n} \Bigl( {\rr \over m_P} \Bigr)^{2n-1} \simeq 0.
\label{approx rad eq}
\end{equation}
Since the kinetic term ${\pi_\rr^2 \over 2 a^3}$ in $H_{\rm eff}$
is much smaller than $ V_{\rm eff} $  for large $\rr$,
by substituting an approximate Friedmann equation
\begin{equation}
{{\dot{a}} \over {a}} \simeq
\sqrt{{{4 \pi m_P^2 \lambda^Q_{2n}} \over {3 n} }
} \Bigl( {\rr \over m_P} \Bigr)^{n},
\end{equation}
into Eq. (\ref{approx rad eq}), we obtain the solutions
\begin{eqnarray}
\rr (t) & \simeq & \rr(t_i)
\exp \Bigl(- \sqrt{{{m_P^2 \lambda^Q_{2n}} \over {6  \pi }} }
(t - t_i) \Bigr), ~ (n =2) ,
\nonumber\\
\rr (t) & \simeq & \rr(t_i)
\Bigl[1 + \sqrt{{{n (n+2)^2
m_P^2 \lambda^Q_{2n}} \over {12  \pi }} }
(t - t_i) \Bigr]^{1/(n-2)}, ~ ( n \geq 3).
\end{eqnarray}
In all cases of $n$ one has a period of inflation described by
\begin{equation}
a(t)  \simeq a (t_i) \exp \Biggl[ \sqrt{{{4 \pi m_P^2 \lambda^Q_{2n}}
\over {3n  }}
} \int_{t_i}^{t} \Bigl({\rr \over m_P} \Bigr)^{n} (t) dt \Biggr],
\end{equation}
that follows from the Friedmann equation. These solutions are the same
as those of the chaotic inflation model \cite{linde,linde3}.
The big difference, however, is that in quantum dynamical
model of inflation the large quantum fluctuation with the nearly minimal uncertainty
and the non-perturbative quantum contributions
at the very early Universe does drive the quasi-exponential
expansion of the Universe. As the Universe inflates,
$\Delta \phi$ decreases but

$\Delta \pi_\rr$ grows and the symmetric
Gaussian state becomes sharply peaked, showing classical features.

Second, we consider the late evolution of the Universe. Using nonlinear
system theory \cite{wiggins}, we find a {\it limit cycle} when
${{\partial} \over {\partial \rr}} V_{eff} (\rr) = 0$:
\begin{equation}
{1 \over 4 {a^6} {\rr^4}} =    m^2 +
m_P^2 \lambda^Q_{2n} \Bigl( {\rr \over m_P} \Bigr)^{2(n-1)} .
\end{equation}
As the Universe expands, the quantum fluctuation
$\rr$ decreases due to the friction
from Hubble parameter and behaves as a classical inflaton field.
The equilibrium is determined dominantly by the mass term,
and the motion of the effective
inflaton tends toward  the limit  cycle of  a
coherent  oscillation  constrained
approximately by
\begin{equation}
\rr^2 a^3 \simeq {{1} \over {2 m}} .
\end{equation}
The Friedmann equation along the limit cycle of the motion
leads to a power-law expansion
\begin{equation}
a (t) \simeq \Bigl({{3 \pi m} \over {2m_P^2}}
 (1 + {{1} \over {n}}) t^2 \Bigr)^{1/3},
\end{equation}
which is valid for the later time.

Finally, we extend the potential in (\ref{e02}) to be an
arbitrary analytic and bounded potential
$V (\phi) = F \left( { {\phi}} \right)$,
which is symmetric, $F(-x) = F(x)$.  One may then
expand the potential in Taylor series by
\begin{equation}
F \left({{\phi} } \right) = \sum_{n = 1}
{m_P^4 C_{2n} \over {(2n)!}} \Bigl({\phi \over m_P} \Bigr)^{2n},
\end{equation}
where the vacuum energy is adjusted to zero. By the same
analysis as before, one may obtain the effective potential
from quantum fluctuation
\begin{equation}
 V_{\rm eff} = {1 \over 8 a^6 \rr^2} + \sum_{n  = 1}
{{m_P^4 C_{2n}} \over {2^{n} n!}}
\Bigl( {\rr^2 \over {m_P^2}} \Bigr)^n .
\end{equation}
Now, most of  the above  analysis of the  nonlinear
system  holds with  a  modification that the
lowest term $n_{\rm min}$ plays the role of
determining the limit cycle and
the largest term $n_{\rm max}$ is driving an inflation.

In summary, we have studied the quantum dynamics of a real inflaton in
its symmetric Gaussian state for a large class  of potentials. We
have found an equivalent two-dimensional nonlinear dynamical system for
the dispersion of the inflaton and the conserved U(1) charge as an
angular momentum. We have shown that the large quantum fluctuation of
the inflaton with a minimal uncertainty naturally explains the
early stage of the inflationary Universe.
It was shown that the U(1) charge is fixed to a definite value,
which plays the role of
angular momentum in the two-dimensional nonlinear system.
This is contrasted to the usual complex inflaton theory
used  in the inflation model
\cite{khalatnikov,scialom} or  the wormhole model \cite{lee}
since there is no a priori reason to take the U(1)
charge to be fixed in these models.

We conclude with some comments that we have not dealt in detail with the
enhancement of effective coupling constants
and with the inhomogeneous degrees of freedom
(i.e. the non-zero modes) in the real scalar field.
Analysis on the non-zero modes may be interesting
since they give rise to the density perturbation
necessary for structure formation
and may affect the inflation and the quantum
dynamics of the real inflaton. The quantization of this interacting theory
of the non-zero modes and the role of effective coupling constants
in density perturbation require a further study.

\bigskip

This work   was supported  in  parts by   Center for Theoretical  Physics,
Seoul   National University, and by the Basic Science Research
Institute Program, Korea Ministry of Education
under Project. No. BSRI-97-2418, BSRI-97-2425 and BSRI-97-2427.
SPK was also supported by Non Directed Research Fund, Korea Research
Foundation, 1997, and JHY by KOSEF Grant No. 97-07-02-02-01-3.

\end{document}